\documentclass[11pt,twoside]{article}

\usepackage{asp2006}
\usepackage{graphicx}

\begin{document}
\title{Determining element abundances of [WC]-type Central Stars for probing stellar evolution and nucleosynthesis}   
\author{H. Todt$^1$, M. Pe\~na$^2$, W.-R. Hamann$^1$, and G. Gr\"afener$^1$}   
\affil{$^1$Universit\"at Potsdam, $^2$UNAM Mexico}    

\begin{abstract} 
[WC]-type CSPNs are hydrogen-deficient Central Stars of Planetary Nebulae
showing strong stellar winds and a carbon-rich chemistry. 
We have analyzed
new high-resolution spectra of [WC]-type CSPNs with the Potsdam Wolf-Rayet
(PoWR) non-LTE expanding atmosphere models, using upgraded model atoms and
atomic data. Previous analyses are repeated on the basis of the current
models which account for iron-line blanketing.  We especially focus on
determining the chemical composition, including some trace elements like
nitrogen which are of key importance for understanding the evolutionary
origin of the hydrogen-deficient Central Stars.
\end{abstract}



\section{Introduction}

Roughly 10\% of the galactic Central Stars of Planetary Nebulae are
hydrogen-deficient
(\citeauthor{Gor01} \citeyear{Gor01}, \citeauthor{Tyl91}\ \citeyear{Tyl91}).
Their
spectra are dominated by helium, carbon, and oxygen 
emission lines and hence very similar to those of massive PopI WC stars. 
Their spectral type is termed [WC], where the brackets distinguish them from
their massive counterparts.

Similar to the PopI WC classification scheme there is a sequence
from  the ``early'' subtypes [WC2-5] showing lines of \mbox{C{\,\sc iv}}, \mbox{He{\,\sc ii}}
and O{\,\sc v-vii} to the ``late'' subtypes [WC6-11] with spectra dominated by
lower ions.  This scheme was later refined by \citet{AckNei03}
such that the earliest types of [WC] stars
which show very strong emission of {O{\,\sc vi}} are designated as [WO1-4] stars. 

It was suggested that this classification scheme corresponds to an
evolutionary sequence from the cooler [WC]-late types to the hotter [WC]-early
types. The determination of element abundances by spectral analyses provides
the empirical base for
understanding the origin and evolution of H-deficient [WC] stars.

\section{Formation of [WC] Central Stars}

According to stellar evolution modeling from \citet{Her01}, 
$[$WC$]$ stars have lost their hydrogen envelope
after the AGB evolution during a last thermal pulse. Depending on the
evolutionary stage at which this pulse overtakes the star there are three possible
scenarios resulting in different element abundances:

The AGB Final Thermal Pulse (AFTP)  at the tip of the AGB only mixes the
hydrogen down, so that e.g. a model for simultaneous burning and mixing
yields mass fractions of  
$X_{\rm H} =0.17 $, $X_{\rm He} = 0.33$, $X_{\rm C} =
0.32$, $X_{\rm O} =0.15 $. 
The Late Thermal Pulse (LTP) occurs just before entering the White Dwarf cooling
track, bringing the star back to the beginning of the post-AGB evolution and
results in $X_{\rm H} =0.02 $, $X_{\rm He} = 0.37$, $X_{\rm C}
=0.40$, $X_{\rm O}=0.18 $.
The last pulse can also happen very late (VLTP), after entering the White Dwarf
cooling track. Then the hydrogen is completely burnt, and enhanced abundances of nitrogen and
neon are predicted. The star is thrown back to the post-AGB phase like in
the LTP scenario and has finally  $X_{\rm He} = 0.31$, $X_{\rm C} = 0.42$,
$X_{\rm O} = 0.23$ and $X_{\rm N} = 0.012$, $X_{\rm Ne} = 0.021$, as derived
by \citet{Alt05}.

\section{Stellar wind models with PoWR}

[WC] stars have extended
atmospheres with  complex ionization structure. For a determination of 
stellar
temperature, mass-loss rate and element abundances by spectral analyses, an appropriate modeling of the wind is necessary.  
Such models are provided by codes like CMFGEN and the Potsdam
Wolf-Rayet model atmospheres (PoWR), which treat the radiative transfer in the
comoving frame in full non-LTE. 
We used our code, PoWR, which includes iron-line blanketing and
clumping. Additionally we implemented higher ions, which were not considered in
previous analyses.

\section{Analysis}

\begin{figure}[ht]
  \vbox{
    \hbox to \hsize{
     \begin{minipage}[b]{0.5\textwidth}
       \center{\includegraphics[width=\textwidth]{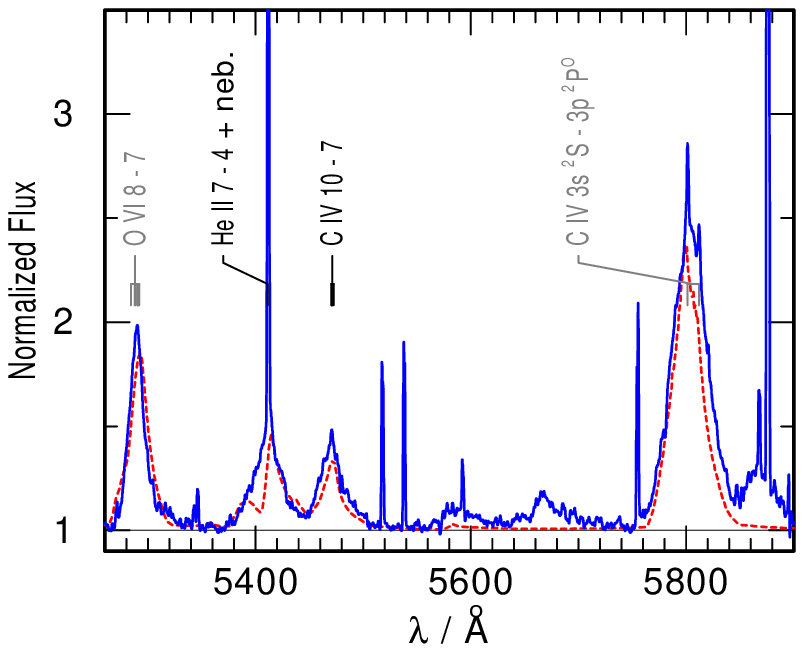}}
     \end{minipage}
     \begin{minipage}[b]{0.47\textwidth}%
       \center{\includegraphics[width=\textwidth]{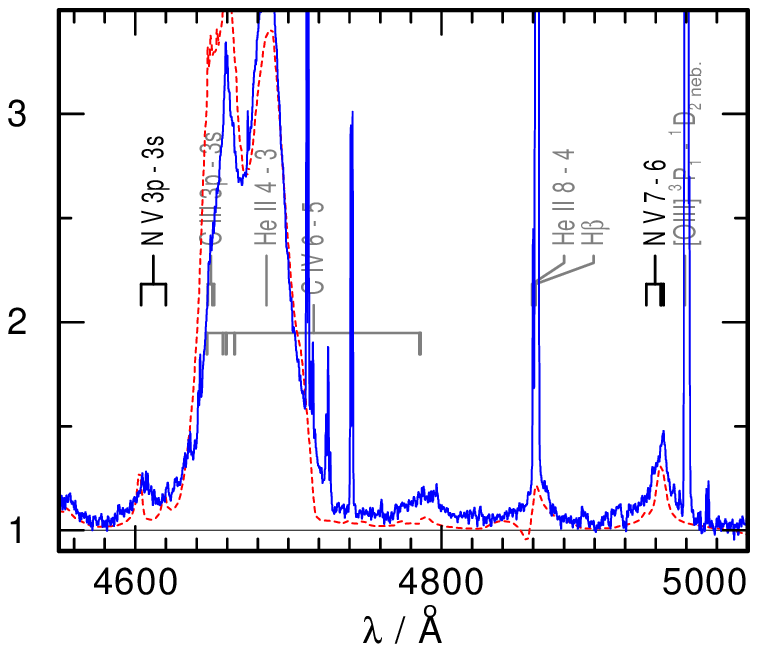}}
     \end{minipage}
    }\vspace{-0.3cm}
    \hbox to \hsize{
      \parbox[t]{0.55\textwidth}{\noindent\caption{Spectrum of
                                 NGC~2867 ([WC2]), observation
                                 (solid line) and model with
                                 $X_{\rm C}$:$X_{\rm He}=$ 26:65 (dashed).
                                 \label{fig:cpnwr12-carbon}}}
      \hfill
      \parbox[t]{0.55\textwidth}{\noindent\caption{Spectrum of PB~6 ([WC2]),
                                 observation (solid) and model with 
                                 $X_{\rm N}=1.5\%$ (dashed). 
                                 \label{fig:pb6-nitrogen}}}
    }
  }
\end{figure}

{\em Carbon and helium.}
Spectral analyses revealed that the main constitutes of
the expanding atmospheres are helium and carbon, in agreement with the
above-mentioned last thermal pulse
scenarios. However, a difference was found between the C:He ratios for the
late-subtypes ([WCL]) and the early-subtypes ([WCE]).  
Previous analyses of [WCL] stars by
Leuenhagen et al. (\citeyear{Leu96}, \citeyear{Leu98}) yielded typical ratios
$X_{\rm C}$:$X_{\rm He}=$ 
50:40. This was also found by \citet{Crow03} and \citet{Ma07}, see also Crowther (these proceedings).
Similar ratios for [WCE] stars were found by \citet{DeMa01} and \citet{Ma07}.
In contradiction, Koesterke \& Hamann (\citeyear{Koe97a}, \citeyear{Koe97b}) found for
[WCE] stars typically $X_{\rm C}$:$X_{\rm He}=$ 30:50, as we confirm by our new analyses.
We employed new high-resolution observations of [WCE] stars
in order to clarify the carbon abundances.
By fitting all carbon and helium lines with special focus on the diagnostic line
pair He{\,\sc ii} (5412\AA) / C{\,\sc iv} (5470\AA), as for NGC~2867 ([WC2]) in
Fig.~\ref{fig:cpnwr12-carbon}, our preference for lower carbon abundance seems
to be confirmed.

{\em Nitrogen.}
Following \citet{WerHer06}, only a VLTP can efficiently produce
nitrogen overabundance.
We focused on two lines of N{\,\sc v} being sensitive
to the nitrogen abundance, namely around 4600 and 4940\,\AA , which are used for
PopI WR  analyses, too. We found $X_{\rm N}= 1\ldots 2\%$ for most of our [WCE] stars, e.g. PB6
(Fig. \ref{fig:pb6-nitrogen}), NGC~2867, and NGC~5189.

{\em Hydrogen.}
Hydrogen could discriminate between the different scenarios, but H lines are
always blended with He{\,\sc ii} lines and nebular emission.  
\citet{Leu98} estimated upper limits for few [WCL] stars and estimated 
$X_{\rm H} =$ 0.01 for PM~1-188 and $X_{\rm H} =$ 0.10 for IRAS~21282+5050.
A more definite detection of $X_{\rm H}  = 0.037$ for V~348~Sgr was reported by \citet{Leu94}.
For the hotter of the [WCE] stars even a mass fraction of $10\%$ would escape detection.

\begin{figure}[b]
\includegraphics[width=\textwidth]{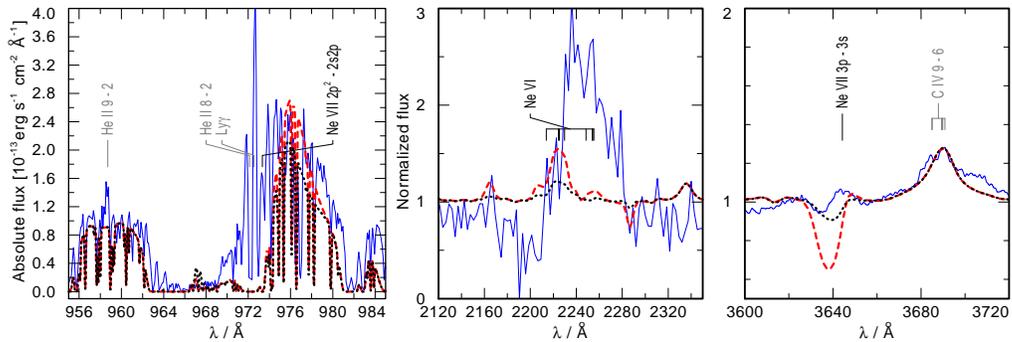}\noindent
\caption{NGC~5189 ([WC2]): observations (solid line) from FUSE, IUE, CTIO 
and models with $X_{\rm Ne}= 3\%$ (dashed) and $0.3\%$
  (dotted) respectively. Models for FUSE range were corrected for ISM
absorption. 
\label{fig:neon-ngc5189}}
\end{figure}

{\em Neon.}
Stars that suffered a VLTP should show an overabundance of neon. \citet{Her05}
found a strong Ne{\,\sc vii} resonance line ({973.33\,\AA}) in 
the FUSE spectrum of NGC~2371. 
Unfortunately this line is almost saturated already at solar neon
abundance, and other lines like the Ne{\,\sc vi} multiplet around {2225\,\AA}
or the Ne{\,\sc vii} absorption line at {3644.3\,\AA} 
are needed for the determination of neon abundances (Fig.~\ref{fig:neon-ngc5189}). 
However, spectra in these ranges are
only available for some of the [WC] stars. 
Furthermore, the modeled line in
the IUE range is neither qualitatively nor quantitatively fitting
the observation with any reasonable neon abundance and is therefore raising
doubts whether the observed P Cygni profile is really due to neon. The line at
{3644.3\,\AA} shows no evidence for higher 
neon abundance.

{\em Iron.}
Iron depletion via s-process nucleosynthesis was predicted by \citet{Her03} and
observed by \citet{StaGrae04} and \citet{Ma07} for [WC] stars.
In contrast, for the [WCE] stars of our analysis we observed
only for NGC~6751 an iron forest but without any unambiguous hint
for a subsolar iron abundance. This is in agreement with models of very hot
[WCE] stars, which do not show an iron forest.  

\section{Conclusions}

The [WCL] analyses by Leuenhagen et
al. (\citeyear{Leu96}, \citeyear{Leu98}) resulted in abundances of neon of $2 \ldots 4\%$,
nitrogen about $1\%$, and carbon around $50\%$, as confirmed by
\citet{DeMa01}, \citet{Crow03} and \citet{Ma07}. 
These abundances would be consistent with the VLTP
scenario. But the detection of hydrogen up to $10\%$ rather points to an AFTP
or LTP origin. 

We re-analyzed [WCE] stars with improved models and new high-resolution spectra
in the optical and FUSE range. The neon lines
could not be fitted consistently, thus leaving the Ne abundance unclear.
More definite is the overabundance of
nitrogen of $1\ldots 2\%$ for many of the [WCE] stars. This points to a
VLTP origin. Contradictory are the systematically lower carbon abundances of
[WC2] stars compared with [WCL] type stars. This challenges the scenario of an
evolutionary
sequence from [WCE] to [WCL]. Moreover, the low carbon abundances are indicative
for the AFTP scenario. But despite problems with analyses one should keep in
mind that even theoretical predictions for carbon abundances scatter.

As the situation is still unclear, we will proceed with our spectral analyses, employing new observations and refining our models with
improved atomic data.

\acknowledgements 
This work was supported by the Bundesministerium f\"ur Bildung und Forschung
(BMBF) under grant 05AVIPB/1.


\end{document}